\documentclass[aps,prl,twocolumn,showpacs,superscriptaddress]{revtex4}
\pdfoutput=1

\usepackage{graphicx}
\usepackage{bm}
\usepackage{amssymb,amsmath,latexsym}
\usepackage{color}
\usepackage{physymb}
\usepackage[colorlinks=true,linktocpage=true,linkcolor=blue,citecolor=blue]{hyperref}

\newcommand{\be}{\begin{equation}}
\newcommand{\ee}{\end{equation}}
\newcommand{\bea}{\begin{eqnarray}}
\newcommand{\eea}{\end{eqnarray}}

\definecolor{darkblue}{RGB}{0,0,196}

\begin{document}

%%%%%%%%%%%%%%%%%%%%%%%%%%%%%%%%%%%%%%%%%%%%%%%%%%%%%%%%%%%%%%%%
% title and abstract
%%%%%%%%%%%%%%%%%%%%%%%%%%%%%%%%%%%%%%%%%%%%%%%%%%%%%%%%%%%%%%%%

\title{Bottomonia suppression in 2.76 TeV Pb-Pb collisions}

\author{Brandon Krouppa}
\affiliation{Department of Physics, Kent State University, Kent, OH 44242 United States}

\author{Radoslaw Ryblewski}
\affiliation {The H. Niewodnicza\'nski Institute of Nuclear Physics, Polish Academy of Sciences, PL-31342 Krak\'ow, Poland}

\author{Michael Strickland}
\affiliation{Department of Physics, Kent State University, Kent, OH 44242 United States}

\begin{abstract}
We compute the QGP suppression of $\Upsilon(1s)$, $\Upsilon(2s)$, $\Upsilon(3s)$, $\chi_{b1}$, and $\chi_{b2}$ states in \mbox{$\sqrt{s_{NN}}=2.76$ TeV} Pb-Pb collisions.  Using the suppression of each of these states, we estimate the inclusive $R_{AA}$ for the $\Upsilon(1s)$ and $\Upsilon(2s)$ states as a function of $N_{\rm part}$, $y$, and $p_T$ including the effect of excited state feed down.  We find that our model provides a reasonable description of preliminary CMS results for the $N_{\rm part}$-, $y$-, and $p_T$-dependence of $R_{AA}$ for both the $\Upsilon(1s)$ and $\Upsilon(2s)$.   Comparing to our previous model predictions, we find a flatter rapidity dependence, thereby reducing some of the tension between our model and ALICE forward-rapidity results for $\Upsilon(1s)$ suppression.
\end{abstract}

\date{\today}

\maketitle

%%%%%%%%%%%%%%%%%%%%%%%%%%%%%%%%%%%%%%%%%%%%%%%%%%%%%%%%%%%%%%%%
% introduction
%%%%%%%%%%%%%%%%%%%%%%%%%%%%%%%%%%%%%%%%%%%%%%%%%%%%%%%%%%%%%%%%

The relativistic heavy-ion collision experiments being carried out at Brookhaven National Laboratory's Relativistic Heavy Ion Collider (RHIC) and CERN's Large Hadron Collider (LHC) study the behavior of matter at extreme temperatures and densities.  The goal of these experiments is to generate a deconfined state of nuclear matter called a quark-gluon plasma (QGP) and to study its properties in detail. Based on hydrodynamic fits to particle production, LHC $\sqrt{s_{NN}} = 2.76$ TeV collisions generate QGP initial temperatures on the order of $T_0 \sim 500-600$ MeV \cite{Heinz:2013th,Gale:2013da}.  At such high temperatures light hadronic states are disassociated and the equation of state of nuclear matter is well-described by a gas of quark and gluon quasiparticles \cite{Haque:2014rua,Mogliacci:2013mca}.  In the transition region between hadronic matter and a proper QGP, the system is composed of liberated quarks and gluons plus a small admixture of heavy bound states.  Although light hadronic states disassociate around the pseudo-critical temperature for the quark-hadron transition, $T_c \sim 165$ MeV, bottomonia, for example, may survive up to temperatures on the order of $T \sim 600\ {\rm MeV} \sim 4 \, T_c$ \cite{Mocsy:2013syh}.  Due to mass/binding-energy ordering of the quarkonium spectrum, one expects that there will be an approximate sequential disassociation, with lighter states ``melting'' before heavier states and excited states melting before their respective ground states \cite{Karsch:2005nk}.   

In this paper, we focus on the suppression of bottomonia in \mbox{$\sqrt{s_{NN}} = 2.76$ TeV} Pb-Pb collisions.   The benefits of working with heavy quarks are that heavy quark bound states are dominated by short distance physics, their binding energies are much smaller than the quark mass $m_Q \gg \Lambda_{\rm QCD}$ ($Q=c,b$), and their sizes are much larger than $1/m_Q$.  As a result, they can be treated using effective field theory methods.  In the heavy quark limit, one finds that a potential-based non-relativistic effective field theory, pNRQCD, can be used to calculate the mass spectrum, decay rates, etc. of heavy quark bound states \cite{Eichten:1979ms,Lucha:1991vn,Brambilla:2004jw,Brambilla:2008cx,Brambilla:2010xn}.  In addition, pNRQCD allows for the systematic inclusion of relativistic corrections.  Using pNRQCD potential models, the vacuum spectrum of all bottomonium states can be reproduced to within less than one percent using a Cornell potential plus spin-spin and spin-orbit interactions \cite{Brambilla:2004wf,Alford:2013jva}.

The use of potential models to describe quarkonium suppression has a long history, starting with the seminal works of Karsch, Matsui, Mehr, and Satz \cite{Matsui:1986dk,Karsch:1987pv} who predicted that quarkonium production would be suppressed in heavy-ion collisions due to Debye-screening in a deconfined QGP.  Using such non-relativistic potential models, there have been studies of quarkonium spectral functions and mesonic current correlators, see e.g. \cite{Mocsy:2004bv,Wong:2004zr,Mocsy:2005qw,Cabrera:2006wh,Alberico:2007rg,Mocsy:2007yj,Rothkopf:2013kya,Burnier:2013nla}.  There have also been lattice QCD calculations of the quarkonium spectral function~\cite{Umeda:2002vr,Asakawa:2003re,Datta:2003ww,Aarts:2007pk,Hatsuda:2006zz,Jakovac:2006sf,Aarts:2010ek,Aarts:2011sm,Burnier:2014ssa,Kim:2014iga}.  Compared to the standard Debye-screened potential models used in early calculations, systematic analysis of the heavy quark potential in the QGP showed that the potential is complex-valued, with the imaginary part of a state's energy being related to the thermal width of the state \cite{Laine:2006ns}.

In the bottom sector, potential model calculations indicate that the $\Upsilon(1s)$, $\Upsilon(2s)$, and $\Upsilon(3s)$ can survive up to temperatures T $\sim$ 593, 228, 172 MeV, respectively \cite{Mocsy:2013syh}.  At these temperatures, the in-medium width of the state becomes on the order of the real part of its binding energy, and the bound state quickly disappears from the spectrum.  However, even below this disassociation point, quarkonia also decay due to in-medium interactions.  For the $\Upsilon(1s)$, the in-medium width approaches 100 MeV at $3\,T_c$~\cite{Margotta:2011ta}.  At this temperature, the $\Upsilon(1s)$ in-medium half-life is on the order 2 \mbox{fm/c}.  Since this is also the timescale over which the QGP evolves hydrodynamically, one needs accurate and reliable modeling of the background evolution in order to make reliable predictions for quarkonium suppression in heavy-ion collisions.

In this paper, we provide an update to the model used in Refs.~\cite{Strickland:2011mw,Strickland:2011aa} to:
(1) extend the background evolution to full (3+1)D anisotropic hydrodynamics (aHydro) with a rapidity profile consistent with experimentally-observed particle multiplicity distributions; (2) update the mixing fractions to recent updated values determined via fits to ATLAS, CMS, and LHCb results for $\Upsilon$ and $\chi_{b}$ production in p-p collisions \cite{Woehri:2014}; (3) correct the probability weight-function used for centrality averaging in order to match the experimental procedure.  We compare the updated model predictions with recently reported results on $\Upsilon$ suppression in Pb-Pb collisions from both the CMS \cite{HIN-10-006} and ALICE \cite{Abelev:2014nua} collaborations.  We find that, with the improvements listed above, the original model of Refs.~\cite{Strickland:2011mw,Strickland:2011aa} gives a reasonable description of the $N_{\rm part}$-, $y$-, and $p_T$-dependence of $\Upsilon(1s)$ and $\Upsilon(2s)$ suppression.

%%%%%%%%%%%%%%%%%%%%%%%%%%%%%%%%%%%%%%%%%%%%%%%%%%%%%%%%%%%%%%%%
% methodology
%%%%%%%%%%%%%%%%%%%%%%%%%%%%%%%%%%%%%%%%%%%%%%%%%%%%%%%%%%%%%%%%

\vspace{2mm}
\noindent {\sl Methodology.} 
For both the potential and dynamical equations used below, we assume that the effective local rest frame (LRF) one-particle distribution function for the particles comprising the QGP is of the form
\be
f(p,x) = f_{\rm eq\!\!}\left(\sqrt{p_T^2 + [1+\xi(x)]p_z^2}\Big/\Lambda(x)\right) ,
\ee 
where $-1 \leq \xi(x) < \infty$ is the local spheroidal momentum-space anisotropy parameter and $\Lambda(x)$ is the local transverse temperature.  This form takes into account the difference between the transverse and longitudinal pressures, which is the most important viscous correction generated in heavy-ion collisions.

As mentioned above, it is now understood that the heavy quark potential in the QGP has both real and imaginary parts, $V = \Re[V] + i \Im[V]$.  We use the internal-energy-based potential specified originally in Ref.~\cite{Strickland:2011aa}.  In the model, the real part of the potential is obtained from the internal energy of the heavy quark/anti-quark system~\footnote{Models based on the free energy seem to be incapable of reproducing either the LHC or RHIC data for $R_{AA}[\Upsilon]$~\cite{Emerick:2011xu,Strickland:2011aa}.}. The resulting real part of the potential is given by \cite{Strickland:2011aa}
\begin{eqnarray}
\Re[V] &=&  -\frac{a}{r} \left(1+\mu \, r\right) e^{-\mu \, r }
+ \frac{2\sigma}{\mu}\left[1-e^{-\mu \, r }\right]
\nonumber \\
&& \hspace{2cm} 
- \sigma \,r\, e^{-\mu \, r } -  \frac{0.8\,\sigma}{m_b^2 r} 
\, , \label{eq:real_pot_model_B}
\end{eqnarray}
where $m_b = 4.7$ GeV, $a=0.385$, $\sigma = 0.223\;{\rm GeV}^2$ \cite{Petreczky:2010yn}, and the last term is a temperature- and spin-independent finite-quark-mass correction taken from Ref.~\cite{Bali:1997am}.  In this expression, $\mu = {\cal G}(\xi,\theta) m_D$ \cite{Dumitru:2007hy,Dumitru:2009ni,Strickland:2011aa} is the anisotropic Debye mass, where ${\cal G}$ is a function which depends on the degree of plasma momentum-space anisotropy $\xi$, the angle of the line connecting the quark-antiquark pair with respect to the beamline  direction $\theta$, and $m_D = 1.4 \sqrt{1+N_f/6} \, g_s T$ is the isotropic leading-order Debye mass adjusted by a factor of $1.4$ in order to take into account higher-order corrections determined via lattice calculations \cite{Kaczmarek:2004gv}.  Note that, in the limit $\xi \rightarrow 0$, one has ${\cal G} = 1$ and the real part of the potential above reduces to the internal energy derived from the original Karsch-Mehr-Satz potential~\cite{Karsch:1987pv}.

The imaginary part of the potential $\Im[V]$ is obtained from a leading-order perturbative calculation performed in the small-$\xi$ limit~\cite{Laine:2006ns,Dumitru:2009fy,Burnier:2009yu}
\begin{eqnarray} 
\Im[V] &=& - \alpha_s C_F T \, \Big\{ \phi(r/m_D) \nonumber \\
&& \hspace{5mm} - \xi \left[ \psi_1(r/m_D,\theta)+\psi_2(r/m_D, \theta)\right] \Big\} \, ,
\label{eq:impot}
\end{eqnarray}
where $\phi$, $\psi_1$, and $\psi_2$ are special functions which can be expressed in terms of the Meijer G-function.  We solve the 3D Schr\"odinger equation with the potential above to obtain the real and imaginary parts of the binding energy as a function of $\xi$ and $\Lambda$ \cite{Margotta:2011ta}.  The imaginary part of the binding energy is then used to obtain the width of each state
\begin{equation}
\Gamma(\tau,{\bf x}_\perp,\varsigma) = 
\left\{
\begin{array}{ll}
2 \Im[E_{\rm bind}]  & \;\;\;\;\; \Re[E_{\rm bind}] >0 \\
\gamma_{\rm dis}  & \;\;\;\;\; \Re[E_{\rm bind}] \le 0 \\
\end{array}
\right. \, ,
\label{eq:width}
\end{equation}
with $\gamma_{\rm dis}$ being the effective decay rate for unbound states, which we take to be 10 GeV~\footnote{We find that our results do not depend in any significant way on $\gamma_{\rm dis}$, as long as $\gamma_{\rm dis} \gtrsim 2$ GeV.}.  It is implicitly understood that $E_{\rm bind}$, and hence $\Gamma$, are local quantities that depend on $\tau = \sqrt{t^2-z^2}$, ${\bf x}_\perp$, and $\varsigma = {\rm tanh}^{-1}(z/t)$ through the (3+1)D evolution of the transverse temperature $\Lambda$, local momentum-space anisotropy $\xi$, and associated flow velocities.  For this purpose, we use (3+1)D anisotropic hydrodynamics (aHydro)~\cite{Florkowski:2010cf,Martinez:2010sc,Strickland:2014pga}.

The (3+1)D aHydro code used provides the spatiotemporal evolution of $\xi$ and $\Lambda$.  The widths obtained from solution of the 3D Schr\"odinger equation are then integrated and exponentiated to compute the relative number of states remaining at a given proper time.  Integrating the instantaneous local decay rate $\Gamma$ over proper-time, one obtains
\begin{eqnarray}
&&R_{\rm AA}(p_T,y,{\bf x}_\perp,b) = e^{-\zeta(p_T,y,{\bf x}_\perp,b)} \nonumber \\
&&\zeta \equiv \Theta(\tau_f-\tau_{\rm form}) \int_{{\rm max}(\tau_{\rm form},\tau_0)}^{\tau_f} 
\! d\tau\,\Gamma(\tau,{\bf x}_\perp,\varsigma=y) \, , \;\;\;\;\;
\label{eq:raa}
\end{eqnarray}
where $b$ is the impact parameter, $\tau_{\rm form} = \gamma \tau_{\rm form}^0 = E_T \tau_{\rm form}^0/M$ where $M$ is the mass of the state, and $\tau_{\rm form}^0$ is the formation time of the state at rest.  For the rest frame formation times, we assume that they are roughly proportional to the inverse vacuum binding energy of each of the states \cite{Karsch:1987uk}.  For the $\Upsilon(1s)$, $\Upsilon(2s)$, $\Upsilon(3s)$, $\chi_{b1}$, and $\chi_{b2}$ states, we use $\tau_{\rm form}^0$ = 0.2, 0.4, 0.6, 0.4, and 0.6 fm/c, respectively~\footnote{We have checked that varying each of these times by $\pm50\%$ results in a $\lesssim 15\%$ variation in our final result for the inclusive $R_{\rm AA}^{\Upsilon(1s)}$.}.

We take the initial proper time $\tau_0$ for hydrodynamic evolution to be $\tau_0 =$ 0.3 fm/c and the initial central temperature for central collisions to be \mbox{$T_0 \in \{ 552, 546, 544 \}$ MeV} for shear viscosity to entropy density ratios $4\pi\eta/s \in \{1,2,3\}$, with the values tuned in order to keep the final charged particle multiplicity fixed.  The final time $\tau_f$ appearing in Eq.~(\ref{eq:raa}) is self-consistently determined from the aHydro simulation as the proper time when local effective temperature becomes less than the transition temperature.  At this effective temperature, plasma screening effects are assumed to decrease rapidly due to the transition to the hadronic phase with the widths of the states becoming approximately equal to their vacuum widths~\footnote{We find that our results are quite insensitive to the final temperature used for the bottomonia integrated decay rate.  This insensitivity is due to the fact that most of the suppression occurs at early times when the temperature is large.}.  For the aHydro initial conditions, we use a smooth linear combination ($\kappa_{\rm binary} = 0.145$) of Glauber wounded-nucleon and binary collision scaling to set the initial energy density profile in the transverse plane.  The inelastic cross-section is taken to be $\sigma_{NN} = 62$ mb.  In the spatial rapidity direction, we use a boost-invariant plateau at central rapidities with Gaussian-tails consistent with limited fragmentation at large rapidity \cite{Bozek:2012qs}
\begin{eqnarray}
f(\varsigma) \equiv \exp \!\left[ - \frac{(\varsigma - \Delta \varsigma)^2}{2 \sigma_\varsigma^2} \Theta (|\varsigma| - \Delta \varsigma) \right] ,
\label{longprof}
\end{eqnarray}
with $\Delta\varsigma = 2.5$ and $\sigma_{\varsigma} = 1.4$ fitted to reproduce the experimental pseudorapidity distribution of charged particles.

In order to compare to the experimental results, we then (a) perform a weighted average over the transverse plane and (b) implement any cuts on centrality, $p_T$, and rapidity necessary.  For the spatial average, the probability distribution function for bottomonium production is taken to be proportional to the local number density of plasma partons $n({\bf x}_\perp,\varsigma)$, i.e. $R_{\rm AA}(p_T,y,b) = (\int_{{\bf x}_\perp} n({\bf x}_\perp,\varsigma) R_{\rm AA}(p_T,y,{\bf x}_\perp))/\int_{{\bf x}_\perp} n({\bf x}_\perp,\varsigma)$.  For implementing $p_T$ cuts, we assume that the $p_T$ probability distribution function is proportional to $E_T^{-4}$.  For implementing cuts in rapidity, we use a flat distribution function.  After implementing the appropriate cuts on $p_T$ and $y$, we obtain $R_{\rm AA}(b)$.  We then convert $b$ to centrality $C$ using the Glauber formalism and integrate over the appropriate centrality cuts using a probability distribution function proportional to $e^{-C/20}$, where $0 < C < 100$.  This probability distribution function takes into account the increased particle production that occurs in central collisions and its form is taken from fits to experimentally observed centrality distributions \cite{Chatrchyan:2012np}.

%%%%%%%%%%%%%%%%%%%%%%%%%%%%%%%%%%%%%%%%%%%%%%%%%%%%
\begin{figure}[t]
\includegraphics[width=0.97\linewidth]{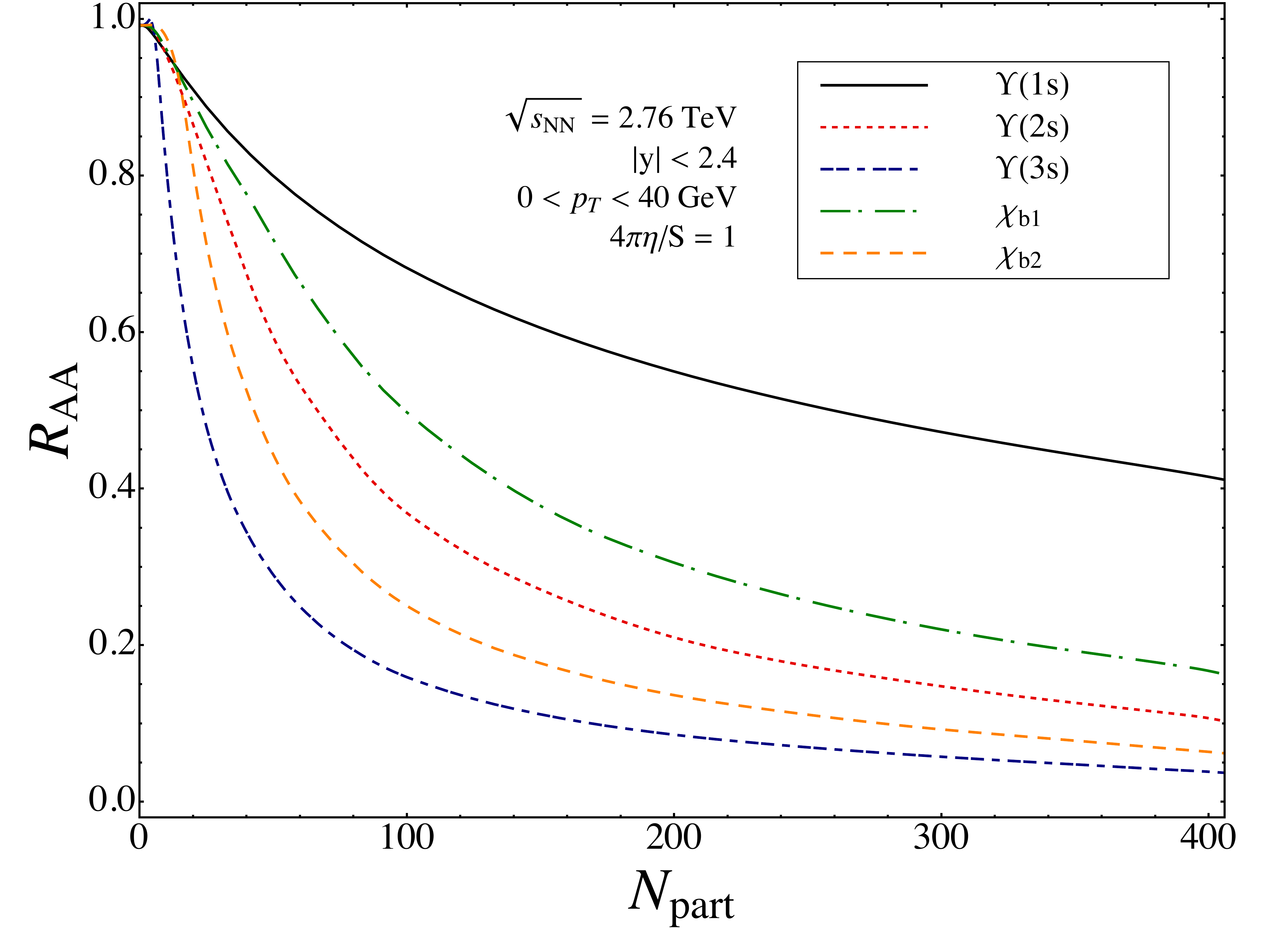}
\vspace{-3mm}
\caption{(Color online) Raw $R_{\rm AA}$ as a function of $N_{\rm part}$.  For this figure we assumed $4\pi\eta/s = 1$.  These curves do not include the effect of excited state feed down. }
\label{fig:raw}
\end{figure}
%%%%%%%%%%%%%%%%%%%%%%%%%%%%%%%%%%%%%%%%%%%%%%%%%%%%

%%%%%%%%%%%%%%%%%%%%%%%%%%%%%%%%%%%%%%%%%%%%%%%%%%%%
\begin{figure}[t]
\includegraphics[width=0.97\linewidth]{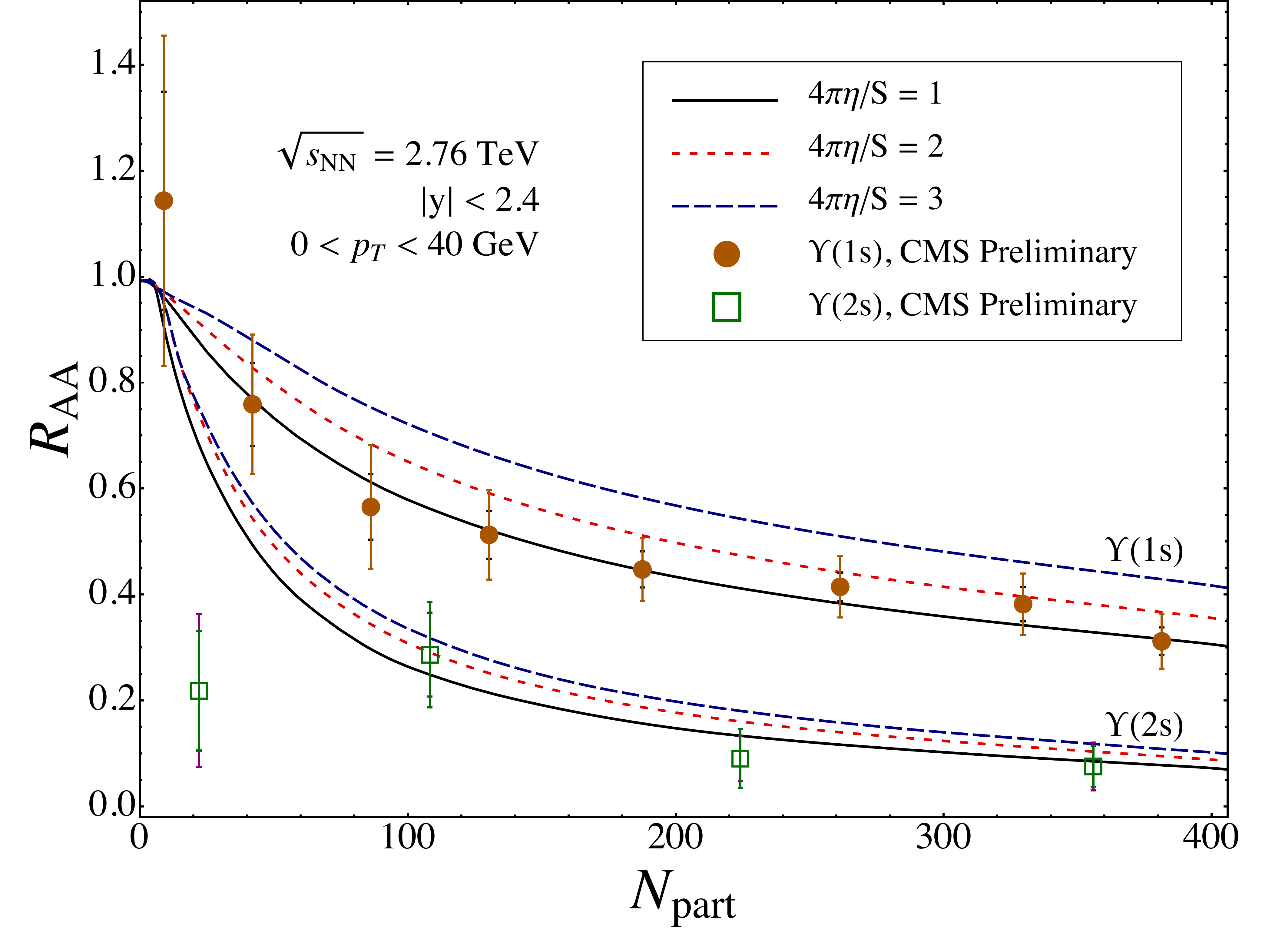}
\vspace{-3mm}
\caption{(Color online) Inclusive $R_{\rm AA}$ for the $\Upsilon(1s)$ and $\Upsilon(2s)$ as a function of $N_{\rm part}$.}
\label{fig:1s2s-npart}
\end{figure}
%%%%%%%%%%%%%%%%%%%%%%%%%%%%%%%%%%%%%%%%%%%%%%%%%%%%

%%%%%%%%%%%%%%%%%%%%%%%%%%%%%%%%%%%%%%%%%%%%%%%%%%%%
\begin{figure}[t]
\includegraphics[width=0.97\linewidth]{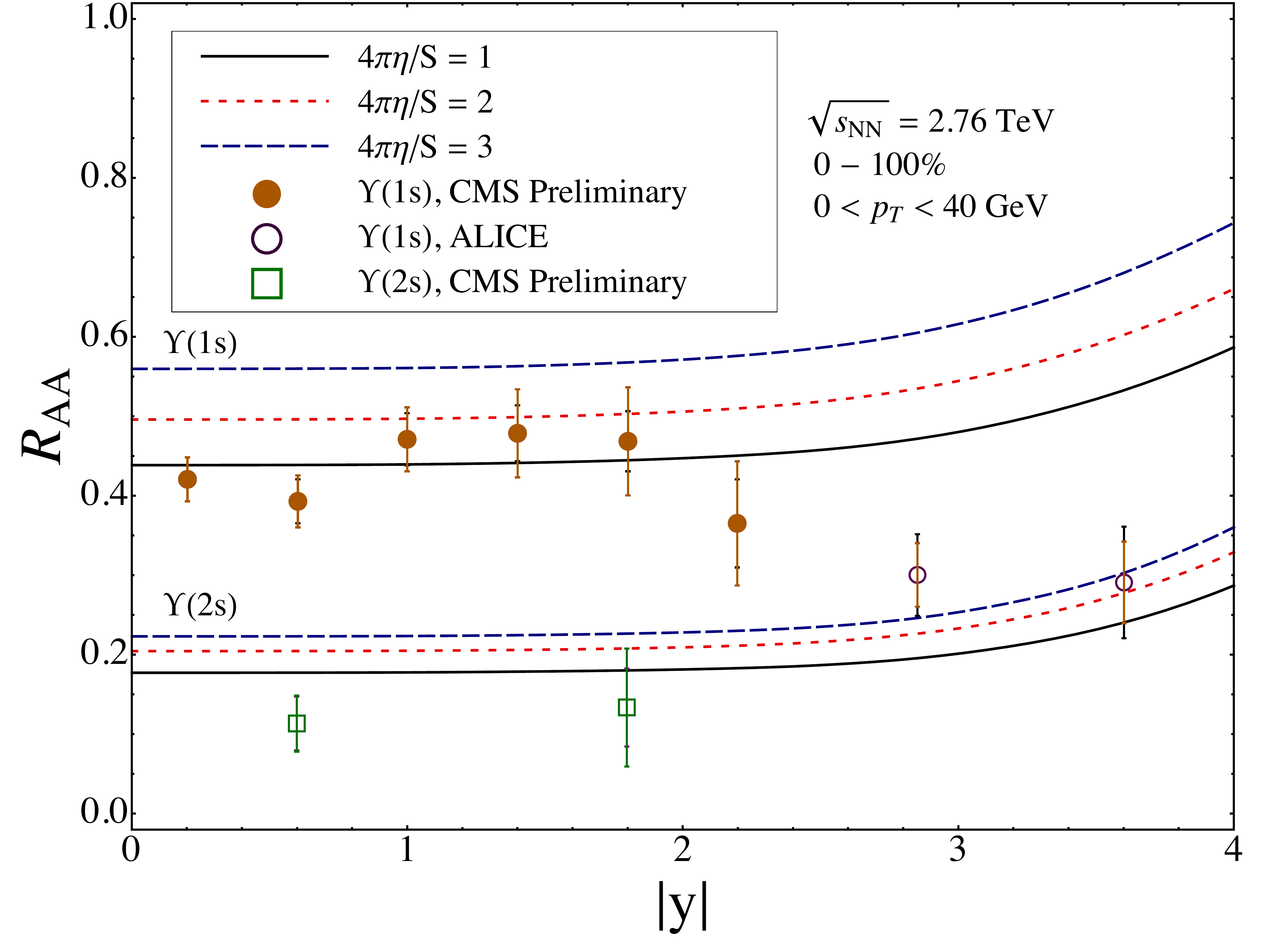}
\vspace{-3mm}
\caption{(Color online) Inclusive $R_{\rm AA}$ for the $\Upsilon(1s)$ and $\Upsilon(2s)$ as a function of $y$.}
\label{fig:1s2s-rap}
\end{figure}
%%%%%%%%%%%%%%%%%%%%%%%%%%%%%%%%%%%%%%%%%%%%%%%%%%%%

%%%%%%%%%%%%%%%%%%%%%%%%%%%%%%%%%%%%%%%%%%%%%%%%%%%%
\begin{figure}[t]
\includegraphics[width=0.97\linewidth]{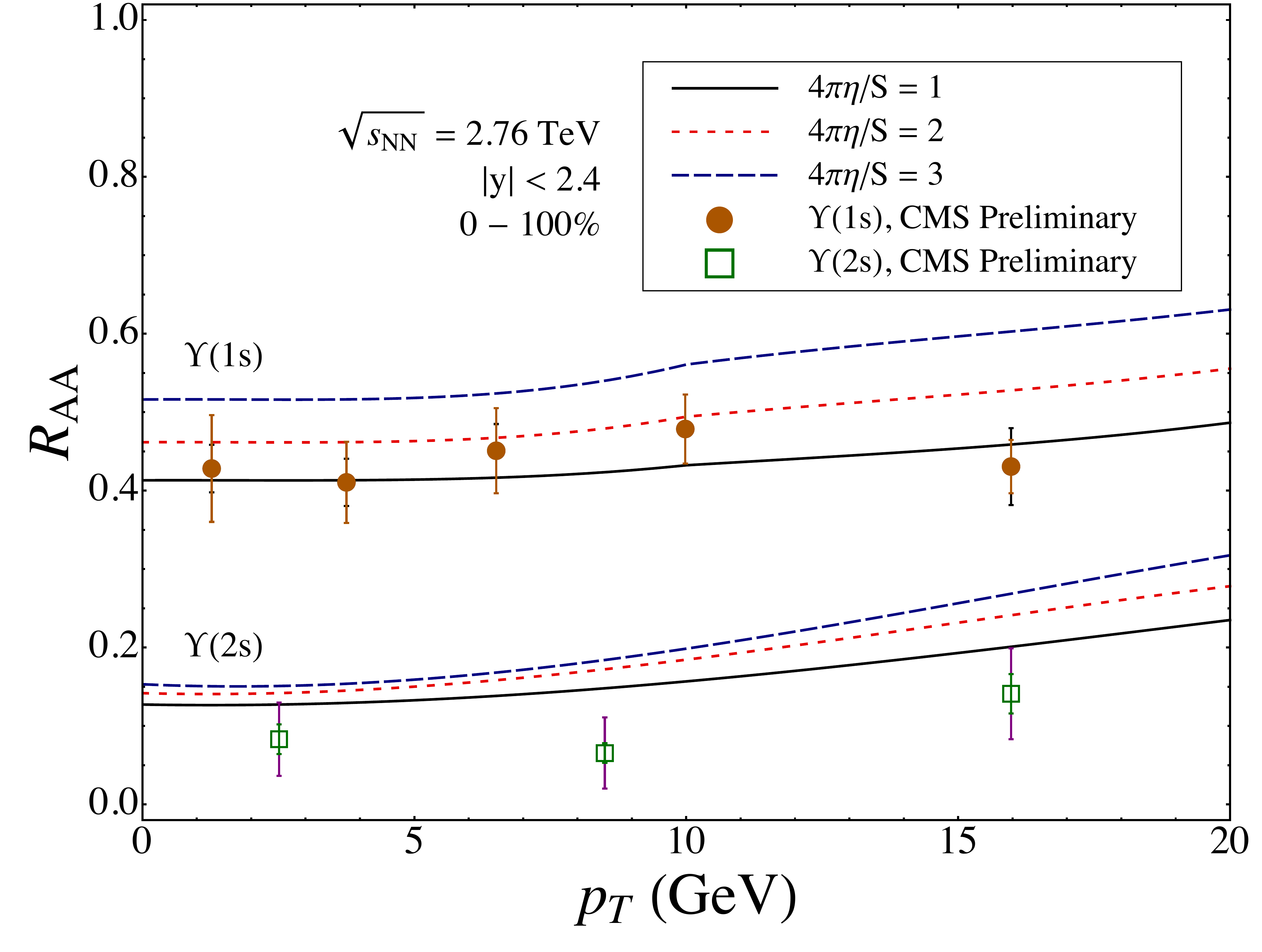}
\vspace{-3mm}
\caption{(Color online) Inclusive $R_{\rm AA}$ for the $\Upsilon(1s)$ and $\Upsilon(2s)$ as a function of $p_T$.}
\label{fig:1s2s-pt}
\end{figure}
%%%%%%%%%%%%%%%%%%%%%%%%%%%%%%%%%%%%%%%%%%%%%%%%%%%%

The procedure outlined above gives the ``raw'' suppression factors for each state.  In order to account for post-QGP feed down of excited states for the $\Upsilon(1s)$, we use $p_T$-averaged feed down fractions obtained recently from a compilation of p-p data available from ATLAS, CMS, and LHCb which gives $f_i^{1s} = \{ 0.618,0.105,0.02,0.207,0.05 \}$ for the $\Upsilon(1s)$, $\Upsilon(2s)$, $\Upsilon(3s)$, $\chi_{b1}$, and $\chi_{b2}$ to $\Upsilon(1s)$ feed down fractions, respectively \cite{Woehri:2014}.  For the $\Upsilon(2s)$, we assume that $f_i^{2s} = \{0.5,0.5\}$ for the $\Upsilon(2s),\Upsilon(3s)$ to $\Upsilon(2s)$ feed down fractions, respectively~\cite{Strickland:2012as}.

%%%%%%%%%%%%%%%%%%%%%%%%%%%%%%%%%%%%%%%%%%%%%%%%%%%%%%%%%%%%%%%%
% Results
%%%%%%%%%%%%%%%%%%%%%%%%%%%%%%%%%%%%%%%%%%%%%%%%%%%%%%%%%%%%%%%%

\vspace{2mm}
\noindent {\sl Results.} 
In Fig.~\ref{fig:raw}, we show the raw $R_{\rm AA}$ for the five states considered as a function of $N_{\rm part}$ for the case that $4\pi\eta/s = 1$.  As can be seen from this figure, there 
is a sequential suppression of the states, however, there are no thresholds visible as originally predicted by sequential suppression~\cite{Karsch:2005nk}.  The lack of thresholds is due to (1) averaging over the full temperature distribution in the transverse plane where the QGP is hotter in the center and colder as one moves towards the edges and (2) the continuous decays of the various states prior to their disassociation point.  Although, we only show results as a function of $N_{\rm part}$, the model provides the full $N_{\rm part}$-, $p_T$-, and $y$-dependence of $R_{\rm AA}$ for each of the states.  

By constructing a linear combination of the raw $R_{\rm AA}$ for each state, we obtain the inclusive $R_{\rm AA}$ for the states.  The result of performing this procedure for the $\Upsilon(1s)$ and $\Upsilon(2s)$ is plotted in Figs.~\ref{fig:1s2s-npart}-\ref{fig:1s2s-pt}.  In these three figures, each set of three lines corresponds to $4\pi\eta/s \in \{1,2,3\}$.  In Fig.~\ref{fig:1s2s-npart}, we compare our results to recently reported preliminary data from the CMS collaboration \cite{HIN-10-006}.  As can be seen from this figure, our model does a good job for both the $\Upsilon(1s)$ and $\Upsilon(2s)$ states.  There is, however, some tension with the lowest $N_{\rm part}$ point for $R_{\rm AA}^{\Upsilon(2s)}$.  Based on the comparison of the model predictions with CMS preliminary data for $R_{\rm AA}^{\Upsilon(1s)}$, the data seem to prefer small shear viscosities in the range $1 \lesssim 4\pi\eta/s \lesssim 2$.  The $R_{\rm AA}^{\Upsilon(2s)}$ data does not seem to provide a tight constraint on $\eta/s$ at this point in time.

In Fig.~\ref{fig:1s2s-rap}, we show our results as a function of rapidity and, once again, we compare with the new CMS preliminary data.  We also include the $R_{\rm AA}^{\Upsilon(1s)}$ result obtained by the ALICE collaboration at forward rapidities as open circles \cite{Abelev:2014nua}.  Although our model does a reasonable job in reproducing the trends seen in the CMS preliminary data, there is still some lingering tension with the ALICE forward results.  We note, however, that compared with earlier predictions made in Ref.~\cite{Strickland:2012as}, our model results are now much closer to the ALICE data.  This is due solely to the change in the way we perform the centrality averaging.  In the past, we used a flat probability distribution as a function of centrality, which does not conform to the procedure used to compute the centrality-averaged results by the experiments, where they simply average over the particles detected in each centrality bin.  With the updated probability distribution function, the centrality-averaged results are much closer to those obtained in central collisions.  

Finally, in Fig.~\ref{fig:1s2s-pt} we show our results as a function of $p_T$ compared to CMS preliminary data.  The flatness of $R_{\rm AA}$ as a function of $p_T$ was a prediction contained in the original model \cite{Strickland:2011mw,Strickland:2011aa} and is due to the fact that, in the model, the bottomonia spectra are assumed to be unaffected due to the lack of thermalization of these states because of their large masses.  The slow increase in $R_{\rm AA}$ as a function of $p_T$ stems solely from the effect of time-dilation of the formation times of the states.  Comparing to the CMS preliminary results for $R_{\rm AA}^{\Upsilon(1s)}$, we see that the data seem to, once again, prefer small values of $\eta/s$.  For the $R_{\rm AA}^{\Upsilon(2s)}$, the model seems to under predict the amount of suppression seen in the CMS preliminary data, however the overall magnitude and weak dependence on $p_T$ predicted by the model seems to be in reasonable agreement with the data.

%%%%%%%%%%%%%%%%%%%%%%%%%%%%%%%%%%%%%%%%%%%%%%%%%%%%%%%%%%%%%%%%
% conclusion
%%%%%%%%%%%%%%%%%%%%%%%%%%%%%%%%%%%%%%%%%%%%%%%%%%%%%%%%%%%%%%%%

\vspace{2mm}
\noindent {\sl Conclusions.} 
In this paper we presented an update to our model predictions for the QGP-induced suppression of bottomonia states at LHC energies \cite{Strickland:2011mw,Strickland:2011aa}.  The potential model itself is exactly the same as used in previously published results, however, we have (1) upgraded the aHydro code to (3+1)D in order to have a more realistic model of the background evolution (2) updated the mixing fractions determined from recent ATLAS, CMS, and LHCb measurements, and (3) corrected our method for performing centrality averaging.  

As can be seen from the results presented herein, the original internal-energy-based model of Refs.~\cite{Strickland:2011mw,Strickland:2011aa} seems to do a reasonable job describing the $N_{\rm part}$-, $y$-, and $p_T$-dependence of CMS preliminary results for $R_{\rm AA}^{\Upsilon(1s)}$ and $R_{\rm AA}^{\Upsilon(2s)}$.  At forward rapidities, there is still some tension with the ALICE $R_{\rm AA}^{\Upsilon(1s)}$ data, however, with the fix to the centrality-averaging procedure, the discrepancy is no longer as dramatic.  Because of this, there is now some hope that the additional suppression at forward rapidities could be explained by cold-nuclear matter effects.  

On the positive side, it seems that for central rapidities ($y \lesssim 2$) the data are consistent with bottomonia suppression due to the creation of a deconfined QGP with a shear viscosity to entropy density ratio roughly between $1/(4\pi)$ and $2/(4\pi)$.  These values are consistent with those obtained via analysis of the collective flow coefficients, thereby providing further evidence that the QGP created in relativistic heavy ion collisions behaves like a nearly perfect fluid.

%%%%%%%%%%%%%%%%%%%%%%%%%%%%%%%%%%%%%%%%%%%%%%%%%%%%%%%%%%%%%%%%
% Acknowledgments
%%%%%%%%%%%%%%%%%%%%%%%%%%%%%%%%%%%%%%%%%%%%%%%%%%%%%%%%%%%%%%%%

\vspace{2mm}
\noindent {\sl Acknowledgments:} We thank N. Filipovic for useful conversations.  M. Strickland and B. Krouppa were supported by the U.S. Department of Energy under Award No.~DE-SC0013470.  R. Ryblewski was supported by the Polish National Science Center Grant No.~DEC-2012/07/D/ST2/02125.

%%%%%%%%%%%%%%%%%%%%%%%%%%%%%%%%%%%%%%%%%%%%%%%%%%%%%%%%%%%%%%%%
% bibliography
%%%%%%%%%%%%%%%%%%%%%%%%%%%%%%%%%%%%%%%%%%%%%%%%%%%%%%%%%%%%%%%%

%==================================================================
\bibliography{bottomonia}
%==================================================================

\end{document}